\documentclass[mypaper,10pt,twoside]{CoAst}
\usepackage{epsf,color,graphicx,picinpar,fancyhdr,multicol,multirow,latexsym,fancybox,amssymb,supertabular,longtable,rotating,nicefrac,amsmath}
\usepackage{lscape} %use as given  \begin{landscape}\begin{table} ... \end{table}\end{landscape}
\usepackage{url} %for proper urls
\usepackage{sfmath}

\usepackage{ucs}
\usepackage{booktabs}
\usepackage{inputenc}
\usepackage{fontenc}
\usepackage{dcolumn}

\usepackage{natbib}

\renewcommand{\chaptermark}[1]{\markboth{#1}{}}

\renewcommand{\headrulewidth}{0pt}
\def\setpubdate{2010}%%%%%%%%%%%%%%%%%%%%%%%%%%%%% pubdate aktualisieren %%%%%%%%%%%%%%%%
\newcommand{\kopf}{\small\itshape 
Comm. in Asteroseismology - Complementary Topics \\ %%%%%%%%%%%% Kopfzeile aktualisieren %%%%%%%%%%%%%%%
Volume 1, \setpubdate \\%%%%%%%%%%%%%%%%%%%%%%%%%%%%%%%%%%%%%%%%%%%%%%%%%%%%%%%%
\copyright~Austrian Academy of Sciences}

\newcommand{\Authors}[1]{\begin{center}\normalsize\bf\sf #1 \end{center}}

\renewcommand{\author}[1]{\begin{center}\normalsize\bf\sf #1 \end{center}}
\newcommand{\Address}[1]{\begin{center}\small\sf #1 \end{center}}

\renewenvironment{abstract}{\section*{Abstract}\normalsize\sf}{}
\newcommand{\References}[1]{\begin{flushleft}{\large References\\}\vspace*{2mm}\small #1 \end{flushleft}}

\newcommand{\chapterCoAst}[2]
{\chapter[\sf\normalsize #1\\ 
\footnotesize \hspace*{5mm}by #2 \sf\normalsize][]
{#1\\}%Produces main paper title
\rhead[\fancyplain{}{\sf\footnotesize \center{#1}}]{\fancyplain{}{\sffamily\thepage}}
\lhead[\fancyplain{\kopf}{\sffamily\thepage}]{\fancyplain{\kopf}{\sf\footnotesize \center{#2}}}}

\newcommand{\figureDSSN}[5]{\begin{figure}[#4]
\centering
\includegraphics*[#5]{#1}
\caption{#2}
\label{#3}
\end{figure}}

\renewcommand{\chaptername}{}

\newcommand{\acknowledgments}[1]{\vspace*{5mm}\noindent  \textbf{Acknowledgments.} #1}

\begin{document}

%\pagenumbering{roman}
\pagestyle{empty}

\sf

%cover.tex----------------------------------------

\setcounter{page}{1}
\thispagestyle{empty}
\vspace*{-1cm}
\begin{center}

%\newpage ~ %%%%%%%%%%%%%%%%%%%%%%%%%%%%%%%%%%%%

%   \textit{Draft Version: \today\\} 
\huge\sf    Communications in Asteroseismology\\
Complementary Topics\\
\vspace*{1cm}
\large 
   Volume 1 \\ \the\year\\
\vspace*{4cm}
\vspace*{4mm}
\vspace*{5cm}

\begin{figure}[!ht]
\centering
\includegraphics*[width=60mm,clip]{OEAW_englisch_2009_SW}

\end{figure}

\normalsize
\end{center}
\newpage
\thispagestyle{empty}
\normalsize
\vspace*{-1cm}

%---Beginn der Impressumsseite
\begin{center}
\begin{large}Communications in Asteroseismology - Complementary Topics\end{large}\\
Editor-in-Chief: \textbf{Michel Breger}, michel.breger@univie.ac.at\\
Editorial Assistant: \textbf{Isolde M\"uller}, isolde.mueller@univie.ac.at\\
Layout \& Production Manager: \textbf{Isolde M\"uller}, isolde.mueller@univie.ac.at\\
%Layout \& Production Assistant: \textbf{Valentina Schmid}, valentina.schmid@univie.ac.at\\
%Language Editor: \textbf{Natalie Sas}, natalie.sas@ster.kuleuven.be\\

\vspace{2mm}
%Institut f\"ur Astronomie der Universit\"at Wien\\
CoAst and CoAct Editorial and Production Office\\
T\"urkenschanzstra\ss e 17, A - 1180 Wien, Austria\\
\textit{http://www.oeaw.ac.at/CoAst/ \\ Comm.Astro@univie.ac.at\\}
\vspace*{4mm}
\textbf{Editorial Board:} Conny Aerts, Gerald Handler, \\ Don Kurtz, Jaymie Matthews, Ennio Poretti\\

\vspace*{1.5cm}

\begin{large}Cover Illustration\end{large}\\
%Simulation of A star surface convections\\ 
%(Illustration kindly provided by F. Kupka and J. Ballot. For more information see the paper by F. Kupka et al., page 30%HÄNDISCH EINGEBEN!!!%%%%%%%%%%%%%%%%%%%%%%%%%%%%%
%)
\end{center}

\vspace*{1.5cm}
\begin{center}
\small
%\vspace{1mm}
\sf British Library Cataloguing in Publication data.\\
\sf A Catalogue record for this book is available from the British Library.
\end{center}
\vspace*{1.4cm}
\small
\begin{center}
All rights reserved\\
ISBN ???-?-????-????-?\\
ISSN ????-????\\
Copyright~\copyright~2010 by\\
Austrian Academy of Sciences \\
Vienna\\
\vspace*{2mm}
Austrian Academy of Sciences Press\\
A-1011 Wien, Postfach 471, Postgasse 7/4\\
Tel. +43-1-515 81/DW 3402-3406, +43-1-512 9050\\ Fax +43-1-515 81/DW 3400\\
http://verlag.oeaw.ac.at, e-mail: verlag@oeaw.ac.at\\
%\normalsize
\end{center}

		\pagestyle{fancyplain}
		\setlength{\columnseprule}{0.2pt}
		\normalsize

		%\vspace*{1cm} %ansonsten auf neuer Seite, d.h. leere Seite wird eingefügt
		%\includepaper{Compendium 2009}{p0_editorial/cia160preface}

%\include{p0_editorial/cia156preface} 
%\newpage

\newpage
\thispagestyle{empty}
\normalsize

\sf
\chapterCoAst{Cinderella User's Manual}{P. Reegen}

\Authors{P. Reegen$^1$} \Address{$^1$ Institut f\"ur Astronomie, T\"urkenschanzstra\ss{}e 17, 1180 Vienna, Austria\\ reegen@astro.univie.ac.at}

\noindent
\begin{abstract}
{\sc Cinderella} is a software solution for the quantitative comparison of time series in the frequency domain. It assigns probabilities to coincident peaks in the DFT amplidude spectra of the datasets under consideration. Two different modes are available. In conditional mode, {\sc Cinderella} examines target and comparison datasets on the assumption that the latter contain artifacts only, returning the conditional probability of a target signal, although there is a coincident signal in the comparison data within the frequency resolution. In composed mode, the probability of coincident signal components in both target and comparison data is evaluated. {\sc Cinderella} permits to examine multiple target and comparison datasets at once.
\end{abstract}

\section{What is {\sc Cinderella}?}

{\sc Cinderella} is an abbreviation of ``{\bf C}omparison of {\bf INDE}pendent {\bf REL}ative {\bf L}east-squares {\bf A}mplitudes''. It provides a quantitative comparison between the DFT amplitude spectra of time-resolved astronomical measurements.

The {\sc SigSpec} technique (Reegen~2005, 2007) allows to determine probabilities for coincident peaks in the DFT amplitude spectra of different datasets quantitatively and in a statistically unbiased way. The theoretical background of this procedure is introduced by Reegen et al.~(2008).

{\sc Cinderella} uses the standard output of the program {\sc SigSpec}, which represents the results of a cascade of consecutive prewhitenings employing least-squares fits to obtain amplitudes and phases of signal components. Following the {\sc SigSpec} terminology, {\sc Cinderella} returns a spectral significances (hereafter abbreviated by `sig') rather than a probability. This manual uses cross-references to {\sc SigSpec} frequently. In these cases, the reader is referred to the {\sc SigSpec} manual (Reegen 2009).

\section{Projects}\label{CINDERELLA_Projects}

{\sc Cinderella} is called by the command line

\begin{scriptsize}\begin{verbatim}
Cinderella <project>
\end{verbatim}\end{scriptsize}

\noindent where {\tt <project>} is the name (or path, if desired) of the {\sc Cinderella} project. The project structure is strictly consistent with {\sc SigSpec}.

Before running the program, the user has to provide

\begin{enumerate}
\item at least two time series input files consistent with the {\sc SigSpec} MultiFile mode (see {\sc SigSpec} manual, p.\,\pageref{SIGSPEC_MultiFile Mode}, and ``Time series input files'', p.\,\pageref{CINDERELLA_Time series input files}), and
\item a directory {\tt <project>} containing the {\sc SigSpec} result files corresponding to the time series input files (see ``{\sc SigSpec} result files'', p.\,\pageref{CINDERELLA_SigSpec result files}).
\end{enumerate}

Furthermore, the user may pass a set of specifications to {\sc Cinderella} by means of a file {\tt <project>.cnd} (see ``The {\tt .cnd} file'', p.\,\pageref{CINDERELLA_The .cnd file}). This file is expected in the same folder as the time series input files and the project directory. For specifications not given by the user, defaults are used.

{\sc Cinderella} is designed to answer two different questions, depending on the problem it is applied to. By default, the program runs through both modes simultaneously and provides both conditional (p.\,\pageref{CINDERELLA_Conditional Mode}) and composed sigs (p.\,\pageref{CINDERELLA_Composed Mode}).

\section{Input}\label{CINDERELLA_Input}

\subsection{Time series input files}\label{CINDERELLA_Time series input files}

{\sc Cinderella} expects at least two time series input files named according to {\tt \#index\#.<project>.dat}. In this context, {\tt \#index\#} is a placeholder for a six-digit index of the file. Note that only files with consecutive indices are appropriately recognised by {\sc Cinderella}. The conventions are compatible with the {\sc SigSpec} MultiFile mode (see {\sc SigSpec} manual, p.\,\pageref{SIGSPEC_MultiFile Mode}), whence the most convenient preparation of data for {\sc Cinderella} is the {\sc SigSpec} MultiFile computation.

The only restrictions to the format of the time series input files are that the number of items per row has to be constant for all rows in the file and that the columns have to be seperated by at least one whitespace character or tab. Dataset entries need not to be numeric, except for the columns specified as time, observable, and weights. The conventions for specifying these three column types are consistent with {\sc SigSpec}. See {\sc SigSpec} manual, pp.\,\pageref{SIGSPEC_Time series columns representing time and observable}, \pageref{SIGSPEC_Time series columns containing statistical weights} for details.

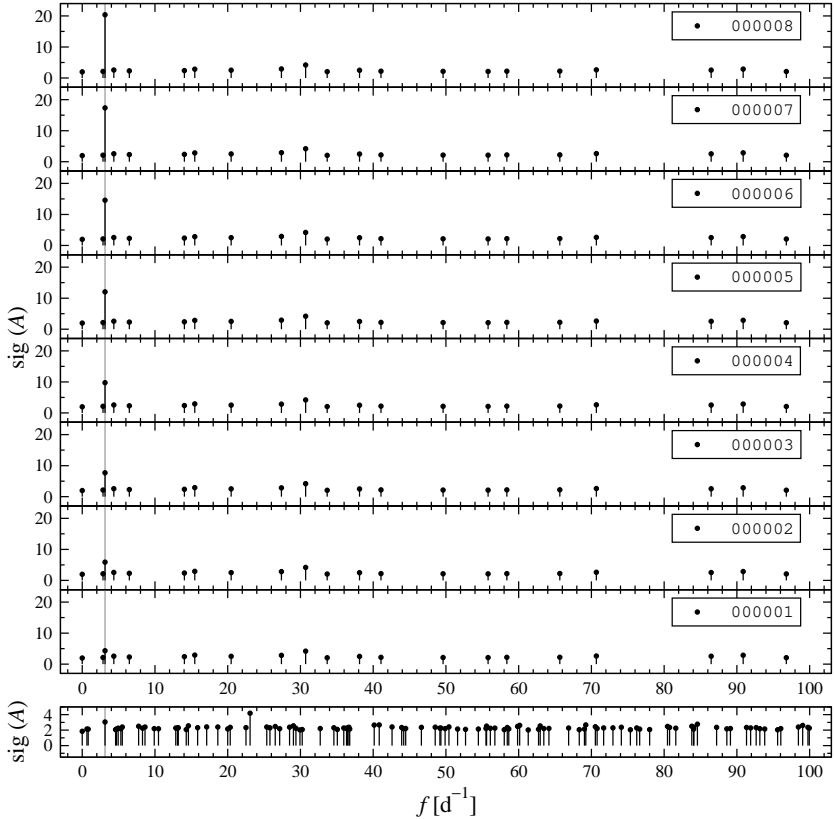
\figureDSSN{f1.eps}{{\sc SigSpec} result files used as input for the sample project {\tt CinderellaNative}. The bottom panel refers to the file {\tt CinderellaNative/000000.result.dat}, which is used as the comparison dataset. Above, the files {\tt CinderellaNative/000001.result.dat} to {\tt CinderellaNative/000008.result.dat} are displayed from bottom to top. The underlying time series represent Gaussian noise plus a single sinusoidal signal at a frequency of $3.125\,\mathrm{d}^{-1}$ (grey line) with different amplitudes.}{CINDERELLA_normalrun.res}{!htb}{clip,angle=0,width=110mm}

\subsection{{\sc SigSpec} result files}\label{CINDERELLA_SigSpec result files}

The {\sc SigSpec} result files {\tt \#index\#.result.dat} are located in the project directory and contain a list of all sig maxima associated to each of the input time series {\tt \#index\#.<project>.dat}. A {\sc SigSpec} result file consists of seven columns. A full description is provided by the {\sc SigSpec} manual, p.\,\pageref{SIGSPEC_Result files}. {\sc Cinderella} uses only five columns:

\begin{itemize}
\item the frequency [inverse time units] (column 1),
\item the sig (column 2),
\item the amplitude [units of observable] (column 3),
\item rms scatter of the time series before prewhitening (column 5), and
\item point-to-point scatter of the time series before prewhitening (column 6).
\end{itemize}

The last line of a result file contains only two non-zero values in columns 5 and 6. These represent the rms and point-to-point scatter of the time series after the last prewhitening step, correspondingly, and are also used by {\sc Cinderella}.

\vspace{12pt}\noindent{\bf Example.}\label{CINDERELLA_EXnormalrun} {\it The sample project {\tt CinderellaNative} provides a run without any additional options by typing {\tt Cinderella CinderellaNative}. The files {\tt 000000.CinderellaNative.dat}, ..., {\tt 000008.CinderellaNative.dat} are the same as {\tt 000038.diffsig.dat} to {\tt 000046.diffsig.dat} in the {\sc SigSpec} example {\tt diffsig} (Reegen 2009, p.\,\pageref{SIGSPEC_EXdiffsig}), correspondingly.

The {\sc SigSpec} result files {\tt 000000.result.dat} to {\tt 000008.result.dat} are provided in the project directory {\tt CinderellaNative}. They contain all signal components found with sig $>$ 2 and are displayed in Fig.\,\ref{CINDERELLA_normalrun.res}\it .

The screen output at runtime starts with a standard header.}

\begin{scriptsize}\begin{verbatim}
 CCCCC  ii            dd                      ll  ll
CC   CC               dd                      ll  ll
CC      ii n nnn   ddddd  eeee  r rrr   eeee  ll  ll   aaaa
CC      ii nn  nn dd  dd ee  ee rr  rr ee  ee ll  ll  aa  aa
CC      ii nn  nn dd  dd ee  ee rr     ee  ee ll  ll      aa
CC      ii nn  nn dd  dd eeeeee rr     eeeeee ll  ll   aaaaa
CC      ii nn  nn dd  dd ee     rr     ee     ll  ll  aa  aa
CC   CC ii nn  nn dd  dd ee  ee rr     ee  ee ll  ll  aa  aa
 CCCCC  ii nn  nn  ddd d  eeee  rr      eeee   ll  ll  aaa a


Comparison of INDEpendent RELative Least-squares Amplitudes
Version 1.0
************************************************************
by Piet Reegen
Institute of Astronomy
University of Vienna
Tuerkenschanzstrasse 17
1180 Vienna, Austria
Release date: April 29, 2008
\end{verbatim}\end{scriptsize}

{\it The program starts with processing the command line, checking if a present directory {\tt CinderellaNative} is present, and searching for a configuration file {\tt CinderellaNative.cnd} (see ``The {\tt .cnd} file'', p.\,\pageref{CINDERELLA_The .cnd file}). Since there is no such file present, {\sc Cinderella} produces a warning message.}

\begin{scriptsize}\begin{verbatim}
*** start **************************************************

Checking availability of project directory CinderellaNative...
project directory CinderellaNative ok.

Warning: CndFile_LoadCnd 001
         Failed to open .cnd file.
\end{verbatim}\end{scriptsize}

{\it Now {\sc Cinderella} counts the time series input files and checks for corresponding {\sc SigSpec} result files.}

\begin{scriptsize}\begin{verbatim}
*** MultiFile count ****************************************

Number of files                           9
\end{verbatim}\end{scriptsize}

{\it The next step is to count the rows in each {\sc SigSpec} result file.}

\begin{scriptsize}\begin{verbatim}
*** count rows in SigSpec result files *********************

CinderellaNative/000000.result.dat:    106 rows
CinderellaNative/000001.result.dat:     21 rows
CinderellaNative/000002.result.dat:     21 rows
CinderellaNative/000003.result.dat:     21 rows
CinderellaNative/000004.result.dat:     21 rows
CinderellaNative/000005.result.dat:     21 rows
CinderellaNative/000006.result.dat:     21 rows
CinderellaNative/000007.result.dat:     21 rows
CinderellaNative/000008.result.dat:     21 rows
\end{verbatim}\end{scriptsize}

{\it Before reading the input files, {\sc Cinderella} performs the assignment of {\tt target}, {\tt comp} and {\tt skip} tags to the datasets (see ``Dataset Types'', p.\,\pageref{CINDERELLA_Dataset Types}). Since there is no file {\tt CinderellaNative.cnd} available, the defaults are used: the first file, {\tt 000000.CinderellaNative.dat} is considered a comparison dataset, the eight remaining files are interpreted as targets.}

\begin{scriptsize}\begin{verbatim}
*** dataset type assignment ********************************


Warning: CndFile_Cind 001
         Failed to open .cnd file.
         Assigning default types.

000000.CinderellaNative.dat: comparison (default)
000001.CinderellaNative.dat: target (default)
000002.CinderellaNative.dat: target (default)
000003.CinderellaNative.dat: target (default)
000004.CinderellaNative.dat: target (default)
000005.CinderellaNative.dat: target (default)
000006.CinderellaNative.dat: target (default)
000007.CinderellaNative.dat: target (default)
000008.CinderellaNative.dat: target (default)

number of target datasets                 8
number of comparison datasets             1
number of datasets to skip                0
\end{verbatim}\end{scriptsize}

{\it The time series and {\sc SigSpec} result files are read, and {\sc Cinderella} displays the frequency resolution and the mean observable for each time series.}

\begin{scriptsize}\begin{verbatim}
*** read input files ***************************************

000000.CinderellaNative.dat: Rayleigh resolution 0.1089880382935977
000000.CinderellaNative.dat: mean observable     0.0074422789800987
CinderellaNative/000000.result.dat
000001.CinderellaNative.dat: Rayleigh resolution 0.0914470160931467
000001.CinderellaNative.dat: mean observable    -0.0704869463068650
CinderellaNative/000001.result.dat
000002.CinderellaNative.dat: Rayleigh resolution 0.0914470160931467
000002.CinderellaNative.dat: mean observable    -0.0875484339175066
CinderellaNative/000002.result.dat
000003.CinderellaNative.dat: Rayleigh resolution 0.0914470160931467
000003.CinderellaNative.dat: mean observable    -0.1046099215281657
CinderellaNative/000003.result.dat
000004.CinderellaNative.dat: Rayleigh resolution 0.0914470160931467
000004.CinderellaNative.dat: mean observable    -0.1216714091388107
CinderellaNative/000004.result.dat
000005.CinderellaNative.dat: Rayleigh resolution 0.0914470160931467
000005.CinderellaNative.dat: mean observable    -0.1387328967494604
CinderellaNative/000005.result.dat
000006.CinderellaNative.dat: Rayleigh resolution 0.0914470160931467
000006.CinderellaNative.dat: mean observable    -0.1557943843601256
CinderellaNative/000006.result.dat
000007.CinderellaNative.dat: Rayleigh resolution 0.0914470160931467
000007.CinderellaNative.dat: mean observable    -0.1728558719707690
CinderellaNative/000007.result.dat
000008.CinderellaNative.dat: Rayleigh resolution 0.0914470160931467
000008.CinderellaNative.dat: mean observable    -0.1899173595814251
CinderellaNative/000008.result.dat
\end{verbatim}\end{scriptsize}

{\it The amplitudes in the {\sc SigSpec} result file of the comparison dataset (file {\tt CinderellaNative/000000.result.dat} have to be transformed in order to be comparable to the target datasets (see ``Amplitude transformation'', p.\,\pageref{CINDERELLA_Amplitude transformation}). {\sc Cinderella} uses the rms residual as a measure for this transformation. This is the default setting.}

\begin{scriptsize}\begin{verbatim}
*** amplitude transformation *******************************

by rms error: file                        0
\end{verbatim}\end{scriptsize}

{\it The core of {\sc Cinderella} consists of three different analyses. The first part of these is the computation of conditional sigs for each pair of target vs.~comparison datasets. In this case we have eight target datasets and one comparison dataset, which results in eight operations. In general, the total number of operations performed in this step is the product of the number of target datasets times the number of comparison datasets. Detailed information on the computation of pairwise conditional sigs is found in ``Single-comparison output'' (p.\,\pageref{CINDERELLA_Single-comparison output}).}\pagebreak

\begin{scriptsize}\begin{verbatim}
*** pairwise Cinderella analysis ***************************

     1 vs.      0: conditional sig
     2 vs.      0: conditional sig
     3 vs.      0: conditional sig
     4 vs.      0: conditional sig
     5 vs.      0: conditional sig
     6 vs.      0: conditional sig
     7 vs.      0: conditional sig
     8 vs.      0: conditional sig
\end{verbatim}\end{scriptsize}

{\it The second part of the {\sc Cinderella} analysis is a computation of mean conditional sigs for each target dataset over all comparison datasets (see ``Multi-comparison output'', p.\,\pageref{CINDERELLA_Multi-comparison output}). Since there is only one comparison dataset available in this example, the output of this procedure is the same as of the previous pairwise analysis. Finally, {\sc Cinderella} evaluates composed sigs for each target dataset. This calculation is applied to the ``raw'' {\sc SigSpec} results and also to the conditional sigs obtained by the previous operation. Details on the computation of composed sigs are found in ``Composed Mode'' (p.\,\pageref{CINDERELLA_Composed Mode}).}

\begin{scriptsize}\begin{verbatim}
*** total Cinderella analysis ******************************

     1: conditional sig
     2: conditional sig
     3: conditional sig
     4: conditional sig
     5: conditional sig
     6: conditional sig
     7: conditional sig
     8: conditional sig
composed sig
composed sig of conditional sigs
\end{verbatim}\end{scriptsize}

{\it On exit, {\sc Cinderella} displays a goodbye message.}

\begin{scriptsize}\begin{verbatim}
Finished.

************************************************************

Thank you for using Cinderella!
Questions or comments?
Please contact Piet Reegen (reegen@astro.univie.ac.at)
Bye!
\end{verbatim}\end{scriptsize}

{\it The {\sc Cinderella} output for this example is discussed in the subsequent chapters.}

\subsection{The {\tt .cnd} file}\label{CINDERELLA_The .cnd file}

An optional file {\tt <project>.cnd} consists of a set of keywords and arguments defining project-specific parameters for {\sc Cinderella}. If this file is not present in the same folder as the time series input files, {\sc Cinderella} uses a set of default parameters.

\vspace{12pt}
{\bf Caution: {\sc Cinderella} demands a carriage-return character at the end of the file {\tt <project>.cnd}, otherwise the program hangs!}
\vspace{12pt}

Lines in the {\tt .cnd} file starting with a {\tt \#} character are ignored by {\sc Cinderella}. This provides the possibility to write comments into the file.

\section{Indexing}\label{CINDERELLA_Indexing}

By default, {\sc Cinderella} expects the file index of time series and {\sc SigSpec} results to start at zero. If the user wants the program to start at a different index, the keyword {\tt mfstart} may be given in the {\tt .cnd} file. This keyword is followed by an integer representing the desired start index. Furthermore, the software takes into account as many files with consecutive indices as available. The number of indices to process may be restricted by means of the keyword {\tt multifile}, followed by an integer representing the last index to take into account. The use of the keywords {\tt mfstart} and {\tt multifile} is strictly consistent with the {\sc SigSpec} conventions ({\sc SigSpec} manual, p.\,\pageref{SIGSPEC_MultiFile Mode}).

\vspace{12pt}\noindent{\bf Example.} \it The sample project {\tt index} contains the same input as the project {\tt CinderellaNative} (p.\,\pageref{CINDERELLA_EXnormalrun}), but both the time series input files and the {\sc SigSpec} result files are now enumerated from {\tt 004847} to {\tt 004855} rather than from {\tt 000000} to {\tt 000008}. The file {\tt index.cnd} consists of a single line,

\begin{scriptsize}\begin{verbatim}
mfstart 4847
\end{verbatim}\end{scriptsize}

\noindent Since the keyword {\tt multifile} is not specified in the file {\tt index.cnd}, {\sc Cinderella} uses all input files available through consecutive indices.

The computations are entirely the same as for {\tt CinderellaNative}, but all indices are consistently incorporated by {\sc Cinderella}. E.\,g., the screen output for the single-comparison computations is now:

\begin{scriptsize}\begin{verbatim}
*** pairwise Cinderella analysis ***************************

  4848 vs.   4847: conditional sig
  4849 vs.   4847: conditional sig
  4850 vs.   4847: conditional sig
  4851 vs.   4847: conditional sig
  4852 vs.   4847: conditional sig
  4853 vs.   4847: conditional sig
  4854 vs.   4847: conditional sig
  4855 vs.   4847: conditional sig
\end{verbatim}\end{scriptsize}\sf

\section{Dataset Types}\label{CINDERELLA_Dataset Types}

In order to avoid unnessecarily high computational effort and redundant output, if comparing all possible pairs of time series input files, {\sc Cinderella} provides the possibility to specify which pairs of target/comparison datasets to take into account. Moreover, the user has the opportunity to identify datasets to be ignored.

{\sc Cinderella} will produce one so-called single-comparison output file (see ``Single-comparison output'', p.\,\pageref{CINDERELLA_Single-comparison output}) for each target-comparison pair. If there is more than one comparison dataset available, additional files are generated for each target. They contain summaries concerning all comparison datasets examined for the target (see ``Multi-comparison output'', p.\,\pageref{CINDERELLA_Multi-comparison output}) and ``Output for composed mode'', p.\,\pageref{CINDERELLA_Output for composed mode}).

Contrary to the file nomenclature, the six-digit format is not required for file indices specified in the {\tt .cnd} file.

\subsection{Target datasets}\label{CINDERELLA_Target datasets}

The keyword {\tt target}\label{CINDERELLA_keyword.target} in the {\tt .cnd} file is used for the specification of a target dataset. The keyword is followed by an integer value referring to the six-digit index of the desired time series input file. Multiple declarations of {\tt target} are supported. If no {\tt .cnd} file is available, {\sc Cinderella} uses the first time series input file (i.\,e. the one with the start index) as the only target dataset. 

\subsection{Comparison datasets}\label{CINDERELLA_Comparison datasets}

The keyword {\tt comp}\label{CINDERELLA_keyword.comp} in the {\tt .cnd} file is used for the specification of a comparison dataset. The keyword is followed by an integer value referring to the six-digit index of the desired time series input file. Contrary to the file nomenclature, the six-digit format is not required for file indices specified in the {\tt .cnd} file. If no {\tt .cnd} file is available, {\sc Cinderella} uses all time series input files -- except for the first one, which is considered target data -- as comparison datasets.

\subsection{Datasets to ignore}\label{CINDERELLA_Datasets to ignore}

The keyword {\tt skip}\label{CINDERELLA_keyword.skip} in the {\tt .cnd} file is used for the specification of a dataset not to be taken into account for computation. The keyword is followed by an integer value referring to the six-digit index of the desired time series input file. Contrary to the file nomenclature, the six-digit format is not required for file indices specified in the {\tt .cnd} file.

\subsection{Default type}\label{CINDERELLA_Default type}

To enhance the convenience for the user, not all files need to be specified by the keywords {\tt target}, {\tt comp} and {\tt skip}. The keyword {\tt deftype}\label{CINDERELLA_keyword.deftype} may be used to assign a default dataset type.
\begin{enumerate}
\item Use {\tt deftype target} to assign the target attribute by default. If no {\tt deftype} keyword is provided, this setting is activated.
\item Use {\tt deftype comp} to assign the comp attribute by default.
\item Use {\tt deftype skip} to assign the skip attribute by default.
\end{enumerate}

\vspace{12pt}\noindent{\bf Example.} \it The sample project {\tt types} contains the same input as the project {\tt CinderellaNative} (p.\,\pageref{CINDERELLA_EXnormalrun}), and the file {\tt types.cnd} contains the two lines

\begin{scriptsize}\begin{verbatim}
deftype target
comp 0
\end{verbatim}\end{scriptsize}

\noindent This reproduces the default assignment of data types, and {\sc Cinderella} performs the same calculations as for the project {\tt CinderellaNative}. The only difference is the screen output:

\begin{scriptsize}\begin{verbatim}
*** dataset type assignment ********************************

000000.types.dat: comparison
000001.types.dat: target
000002.types.dat: target
000003.types.dat: target
000004.types.dat: target
000005.types.dat: target
000006.types.dat: target
000007.types.dat: target
000008.types.dat: target
\end{verbatim}\end{scriptsize}

\noindent The fact that {\tt (default)} is not attached to the file list indicates that {\sc Cinderella} uses the specifications given in the file {\tt types.cnd} rather than the standard settings applied to the project {\tt CinderellaNative}.\sf

\section{Conditional Mode}\label{CINDERELLA_Conditional Mode}

The conditional sig is a measure of the probability that a signal component in a target star occurs, although a coincident signal component is found in a comparison star or sky background. It provides an answer to the question, ``What is the probability that a signal component with given amplitude and sig in the target data is not due to the same process that produces a coincident signal component with given amplitude and sig in the comparison data?''

The conditional {\sc Cinderella} mode is comparable to the differential sig ({\sc SigSpec} manual, p.\,\pageref{SIGSPEC_Differential significance spectra}), although the numerical results are not the same.

\begin{itemize}
\item For the differential mode of {\sc SigSpec}, the full spectral information is available. Thus {\sc SigSpec} handles the DFT spectra as continuous functions. {\sc Cinderella} accesses only a list of peaks detected by {\sc SigSpec}. Deviations of corresponding peak frequencies in comparison and target spectra cannot be handled as accurately as in the case of differential sig computation.
\item The differential mode of {\sc SigSpec} compares power integrals over the entire frequency range under consideration for the transformation of amplitudes from comparison into target data. Since {\sc Cinderella} deals with a list of frequencies rather than the entire spectra, different strategies to transform amplitudes have to be employed. See ``Amplitude transformation'', p.\,\pageref{CINDERELLA_Amplitude transformation}.
\end{itemize}

\subsubsection{Single-comparison output}\label{CINDERELLA_Single-comparison output}

The single-comparison output files contain three columns:

\begin{enumerate}
\item target frequency [inverse time units],
\item conditional sig,
\item conditional csig.
\end{enumerate}

Each file refers to the analysis of a single target-comparison dataset pair. Correspondingly, two six-digit indices {\tt \#target\#}, {\tt \#comparison\#} are used to form the filenames, {\tt \#target\#.cd.\#comparison\#.dat} (conditional mode) and {\tt \#target\#.cd.\#comparison\#.dat} (composed mode).

\vspace{12pt}\noindent{\bf Example.} {\it The output file {\tt 000004.cd.000026.dat} contains conditional sigs for {\tt 000004.<project>.dat} as target data and {\tt 000026.<project>.dat} as comparison data.}

\vspace{12pt}\noindent{\bf Example.} \it In the sample {\tt CinderellaNative}, there are 8 single-compari\-son output files: {\tt 000001.cd.000000.dat} to {\tt 000008.cd.000000.dat}. Since there is only one comparison dataset available, these files are redundant, because {\tt \#target\#.cd.dat} $=$ {\tt \#target\#.cd.000000.dat}.\sf

\subsubsection{Multi-comparison output}\label{CINDERELLA_Multi-comparison output}

For each target dataset, a multi-comparison output file {\tt \#target\#.cd.dat} is generated for each target dataset. The three columns represent

\begin{enumerate}
\item target frequency [inverse time units],
\item mean conditional sig for all comparison datasets,
\item mean conditional csig for all comparison datasets.
\end{enumerate}

If there is only one comparison dataset available, the single-comparison and multi-comparison output files are identical.

\vspace{12pt}\noindent{\bf Example.} \it The sample {\tt CinderellaNative} contains 8 multi-comparison output files in the project directory: {\tt 000001.cd.dat} to {\tt 000008.cd.dat}.\sf

\subsection{Candidate selection}\label{CINDERELLA_Candidate selection}

For each target dataset, {\sc Cinderella} scans all comparison datasets, searching for coincident signal components. A pair of signal components is considered coincident, if the frequencies match to an accuracy that may be specified by the user, who may also define what {\sc Cinderella} shall do if a comparison dataset does not contain a match for a given target frequency.

If more than one coincident frequency associated to a given target frequency is found, {\sc Cinderella} chooses the candidate with the lowest conditional or composed sig in order to obtain the most conservative solution.

\subsubsection{Frequency resolution}\label{CINDERELLA_Frequency resolution}

There are mainly two interpretations of the frequency resolution. It is either calculated as the inverse time interval width (Rayleigh frequency resolution),
\begin{equation}\label{CINDERELLA_EQ Rayleigh}
\delta f := \frac{1}{T}\: ,
\end{equation}
or as
\begin{equation}\label{CINDERELLA_EQ Kallinger}
\delta f := \frac{1}{T\sqrt{\mathrm{sig}\left( A\right)}}\: ,
\end{equation}
where $\mathrm{sig}\left( A\right)$ denotes the sig of an amplitude $A$. This definition is called Kallinger resolution (Kallinger, Reegen \& Weiss 2008).

In order to enhance the flexibility of {\sc Cinderella}, the frequency resolution is computed as \begin{equation}\label{CINDERELLA_EQ fres}
\delta f := \frac{1}{T\sqrt{\mathrm{sig}\left( A\right)^{\tau}}}\: ,
\end{equation}
where the floating-point number $\tau\in\left[ 0,1\right]$ may be defined using the keyword {\tt tol}\label{CINDERELLA_keyword.tol} in the {\tt .cnd} file. The special value $\tau = 0$ transforms eq.\,\ref{CINDERELLA_EQ fres} into eq.\,\ref{CINDERELLA_EQ Rayleigh}, setting $\tau = 1$ provides eq.\,\ref{CINDERELLA_EQ Kallinger}.
The default value is $\tau = 0$.

{\sc Cinderella} checks for frequencies in the comparison datasets are within the frequency resolution around each frequency in the target dataset.

\vspace{12pt}\noindent{\bf Example.} \it The sample {\tt cand} contains the same input as {\tt CinderellaNative} (p.\,\pageref{CINDERELLA_EXnormalrun}), and the file {\tt cand.cnd} contains the line

\begin{scriptsize}\begin{verbatim}
tol 2
\end{verbatim}\end{scriptsize}

\noindent The frequency tolerance parameter is increased compared to the default value 0, which means that the intervals taken into account to search for corresponding signal components are tendentially narrower. This setting is for demonstration only; in normal applications, only frequency tolerance parameters ranging from 0 to 1 will make sense.

The effect of this modification is visible, e.\,g., comparing the output files {\tt 000001.cd.000000.dat} of the project {\tt CinderellaNative} to the project {\tt cand}. In the project {\tt CinderellaNative}, this file contains the line

\begin{scriptsize}\begin{verbatim}
58.3815412948909298 -8.8063413553403507 -5.6290446018702553
\end{verbatim}\end{scriptsize}

\noindent whereas the corresponding line in the project {\tt cand} is

\begin{scriptsize}\begin{verbatim}
58.3815412948909298       2.1858505500938747       0.3972034177439077
\end{verbatim}\end{scriptsize}

\noindent In the project {\tt CinderellaNative}, the 14th component in the {\sc SigSpec} result file {\tt 000001.result.dat} in the project directory is related to the 47th component in the file {\tt 000000.result.dat}. The two frequencies differ by 0.054, and the Rayleigh frequency resolution of the target dataset is $0.091$, which is sufficient for a correspondence. In the project {\tt cand}, the target sig of 2.387 becomes relevant. Eq.\,\ref{CINDERELLA_EQ fres} \it yields a frequency resolution 0.042 for this component, which is now too small for a coincidence. The corresponding line in the file {\tt 000001.cd.000000.dat} consistently indicates that no coincident peak is found for this signal component. In this case, {\sc Cinderella} uses a default sig threshold for the comparison data, see ``Spectral significance threshold'', below.\sf

\subsubsection{Spectral significance threshold}\label{CINDERELLA_Spectral significance threshold}

If no coincidence in a comparison dataset is detected for a given target frequency, i.\,e., if {\sc Cinderella} does not find a valid candidate for this target frequency, a default value is used for the sig in the comparison dataset. The user may specify this {\sc Cinderella} threshold by means of the keyword {\tt defsig}\label{CINDERELLA_keyword.defsig} in the {\tt .cnd} file. The same specification may be set for the default csig\footnote{abbreviation for cumulative sig} using the keyword {\tt defcsig}\label{CINDERELLA_keyword.defcsig}. If one of these keywords is not provided, $\frac{\pi}{4}\log\mathrm{e} \approx 0.341$ is used correspondingly by default. According to Reegen~(2007), this is the expected value of the sig for white noise. The underlying assumption is that the residuals after prewhitening of all significant signal components in the comparison dataset represent pure noise, i.\,e.~do not contain any further unresolved signal.

\vspace{12pt}\noindent{\bf Example.} \it The sample {\tt cand} contains the same input as {\tt CinderellaNative} (p.\,\pageref{CINDERELLA_EXnormalrun}), and the file {\tt cand.cnd} contains the two lines

\begin{scriptsize}\begin{verbatim}
defsig 0
defcsig 1
\end{verbatim}\end{scriptsize}

The second row in the output file {\tt 000001.cd.000000.dat} of the project {\tt CinderellaNative}

\begin{scriptsize}\begin{verbatim}
30.7091991449662061 3.8506295783390758 3.6090130812823817
\end{verbatim}\end{scriptsize}

\noindent using the default sig and csig threshold $\frac{\pi}{4}\log\mathrm{e}$, because the comparison dataset does not contain a coincident signal component. The second row in the corresponding file of the project {\tt cand} is

\begin{scriptsize}\begin{verbatim}
30.7091991449662061 4.1917236667995361 2.9501071697428420
\end{verbatim}\end{scriptsize}

\noindent Since the default sig is lower in the project {\tt cand}, the resulting conditional sig is higher. On the other hand, the default csig is higher, which causes the resulting conditional csig to drop down.\sf

\subsection{Amplitude transformation}\label{CINDERELLA_Amplitude transformation}

The assumption that instrumental and environmental artifacts use to be additive in terms of intensity may create needs to transform amplitudes in mag from the comparison into the target spectra, if the conditional {\sc Cinderella} mode is applied. The amplitude transformation is only performed to obtain conditional sigs.

Three different strategies to adjust comparison amplitudes are offered, according to the specifications in the {\tt .cnd} file.

\begin{itemize}
\item In photometry, the photon statistics may be employed to transform comparison into target amplitudes, if the mean magnitudes of the stars are known (Reegen et al.~2008). If the keyword {\tt transam:mean}\label{CINDERELLA_keyword.transam:mean} is provided in the {\tt .cnd} file, {\sc Cinderella} uses the mean observables $\left< m_C\right>$, $\left< m_T\right>$ of the comparison and target time series, respectively, to transform the comparison amplitude $A_C$ into a target amplitude $A_T$ according to
\begin{equation}\label{CINDERELLA_EQtransam_mean}
A_T = 2.5\log\left[ 1 + \frac{10^{-0.4\left(\left< m_C\right> - A_C\right)} - 10^{-0.4\left< m_C\right>}}{10^{-0.4\left< m_T\right>}}\right]\: .
\end{equation}
\item If the keyword {\tt transam:rms}\label{CINDERELLA_keyword.transam:rms} is specified in the {\tt .cnd} file, {\sc Cinderella} interprets the residual rms errors $\sigma _C$, $\sigma _T$ of the comparison and target time series\footnote{with all significant signal prewhitened}, respectively, as measures of the photon noise levels and evaluates the transformed amplitude according to
\begin{equation}\label{CINDERELLA_EQtransam_rms}
A_T = 2.5\log\left[ 1 + \frac{\sigma _C^2}{\sigma _T^2}\left( 10^{0.4\,A_C} - 1\right)\right]\: .
\end{equation}
\item The keyword {\tt transam:ppsc}\label{CINDERELLA_keyword.transam:ppsc} in the {\tt .cnd} file causes {\sc Cinderella} to use Eq.\,\ref{CINDERELLA_EQtransam_rms} employing residual point-to-point scatter instead of residual rms error for both $\sigma _C$ and $\sigma _T$.
\item If the keyword {\tt transam:none}\label{CINDERELLA_keyword.transam:none} is specified in the {\tt .cnd} file, no amplitude transformation is performed at all, i.\,e., {\sc Cinderella} assumes $A_T = A_C$.
\end{itemize}

\vspace{12pt}\noindent{\bf Example.} \it The sample project {\tt transam-mean} contains the same input as the project {\tt CinderellaNative} (p.\,\pageref{CINDERELLA_EXnormalrun}). The time series data are considered to represent millimag photometry. The comparison dataset is assumed to refer to a 5 mag star, whereas the target datasets shall correspond to a 15 mag star. The resulting time series input files are {\tt 000000.transam-mean.dat} {\tt 000001.transam-mean.dat} to {\tt 000008.transam-mean.dat}. The keyword

\begin{scriptsize}\begin{verbatim}
transam:mean
\end{verbatim}\end{scriptsize}

\noindent in the file {\tt transam-mean.cnd} forces {\sc Cinderella} to employ the mean magnitudes of the datasets for the amplitude transformation.\sf

\vspace{12pt}\noindent{\bf Example.} \it The sample project {\tt CinderellaNative} contains an amplitude transformation based on the rms residual, which is the default method.\sf

\vspace{12pt}\noindent{\bf Example.} \it The sample project {\tt transam-ppsc} contains the same input as {\tt CinderellaNative} (p.\,\pageref{CINDERELLA_EXnormalrun}). The line

\begin{scriptsize}\begin{verbatim}
transam:ppsc
\end{verbatim}\end{scriptsize}

in the file {\tt transam-ppsc.cnd} forces {\sc Cinderella} to employ the residual point-to-point scatters of the datasets for the amplitude transformation.\sf

\vspace{12pt}\noindent{\bf Example.} \it The sample project {\tt transam-none} contains the same input as {\tt CinderellaNative} (p.\,\pageref{CINDERELLA_EXnormalrun}). The line

\begin{scriptsize}\begin{verbatim}
transam:none
\end{verbatim}\end{scriptsize}

\noindent in the file {\tt transam-none.cnd} switches off the amplitude transformation.\sf

\section{Composed Mode}\label{CINDERELLA_Composed Mode}

The composed sig is a measure of the probability that two coincident signal components occur in two different datasets. This implements a logical `and', providing an answer to the question, ``What is the probability that two different datasets show coincident signal components with given amplitudes and sigs?''

The composed mode is useful for, e.\,g., photometry of the same star in different filters, or if two short datasets of the same object obtained in different years are examined.

Note that the composed sig in the {\sc SigSpec} result files (see {\sc SigSpec} manual) is consistently defined, but applies to the set of significant signal components displayed in the file, whereas {\sc Cinderella} refers to the composed sig of signal components found in two or more different datasets.

Contrary to the candidate selection procedure in conditional mode (p.\,\pageref{CINDERELLA_Candidate selection}), the frequency interval between the lowest and the highest frequency found in all target datasets is scanned in steps defined by half the frequency resolution (p.\,\pageref{CINDERELLA_Frequency resolution}). For each of the frequencies under consideration, {\sc Cinderella} computes a composed sig, basically following the introduction by Reegen et al.~(2008). Since {\sc Cinderella}'s composed mode takes into account all signal components in all datasets, statistical weights have to be introduced that put more emphasis to signal frequencies closer to the frequency under consideration. Hence the composed sig $\mathrm{csig}\left( A_n\right)$ (annotation by Reegen et al.~2008) assigned to an arbitrary frequency $f$ is evaluated according to
\begin{equation}
\log\left[ 1-10^{\mathrm{csig}\left( A_n\right)}\right] = \frac{1}{N}\sum _{n=1}^{N}\mathrm{e} ^{-\frac{1}{2}\left[\frac{f-f_n}{\min\left(\delta f_n\right)}\right]^2}\log\left[ 1 - 10^{-\mathrm{sig}\,\left( A_n\right)}\right]\: .
\end{equation}
In this context, the total number of signal components in all target datasets is denoted $N$, $f_n$ referring to the frequency of one of these signal components. The minimum frequency resolution $\min\left(\delta f_n\right)$ incorporates the definition by Eq.\,\ref{CINDERELLA_EQ fres}.

An interpolation loop is used to exactly identify the maxima of this composed sig, which are written to the output file.

\subsection{Output for composed mode}\label{CINDERELLA_Output for composed mode}

The calculation of the composed sig is applied to all target datasets at once. A file {\tt cp.dat} is generated, which contains the composed sigs of all target datasets. The three columns refer to:

\begin{enumerate}
\item target frequency [inverse time units],
\item composed sig for all target datasets,
\item composed csig for all target datasets.
\end{enumerate}

The composed sigs are also calculated for the conditional sigs, i.\,e., the files {\tt \#target\#.cd.dat} (see ``Multi-comparison output'', p.\,\pageref{CINDERELLA_Single-comparison output}), and written to a file {\tt cc.dat}. The column format is the same as for the file {\tt cp.dat}.

\vspace{12pt}\noindent{\bf Example.} \it For the sample project {\tt CinderellaNative}, the project directory contains a file {\tt cp.dat} with the composed sigs (and csigs) for all target datasets (provided by the {\sc SigSpec} result files {\tt 000001.result.dat} to {\tt 000008.result.dat}). Furthermore, a file {\tt cc.dat} is found in the project directory. It contains the composed sigs (and csigs) applied to the conditional ones for all target datasets, i.\,e., the composed sigs are evaluated using the multi-comparison output files generated by the conditional mode, {\tt 000001.cd.dat} to {\tt 000008.cd.dat}.\sf

\section{Keywords Reference}\label{CINDERELLA_Keywords Reference}

This section is a compilation of all keywords accepted by {\sc Cinderella}. A brief description of arguments and default values is given. The type of argument is provided by either {\tt <int>}, {\tt <double>}, or {\tt <string>}, and default values are given in parentheses, e.\,g.~{\tt (2)}. Empty parentheses indicate that there is no default setting.

\subsubsection{\tt col:obs <int> (2)}

observable column index (unique), starting with $1$, {\sc SigSpec} manual, p.\,\pageref{SIGSPEC_keyword.col:obs}

\subsubsection{\tt col:time <int> (1)}

time column index (unique), starting with $1$, {\sc SigSpec} manual, p.\,\pageref{SIGSPEC_keyword.col:time}

\subsubsection{\tt col:weights <int>}

weights column index (also multiple), starting with $1$, {\sc SigSpec} manual, p.\,\pageref{SIGSPEC_keyword.col:weights}

\subsubsection{\tt comp <int> (all except start index)}

specification of time series input files to be regarded as comparison datasets, p.\,\pageref{CINDERELLA_keyword.comp}

\subsubsection{\tt defcsig <double> ($\frac{\pi}{4}\log\mathrm{e} \approx 0.341$)}

threshold to be used for the csig, if no coincidence is found in a comparison dataset, p.\,\pageref{CINDERELLA_keyword.defcsig}

\subsubsection{\tt defsig <double> ($\frac{\pi}{4}\log\mathrm{e} \approx 0.341$)}

threshold to be used for the sig, if no coincidence is found in a comparison dataset, p.\,\pageref{CINDERELLA_keyword.defsig}

\subsubsection{\tt deftype <target/comp/skip> (target)}

specifies the type of dataset to be assigned to a time series by default, p.\,\pageref{CINDERELLA_keyword.deftype}

\subsubsection{\tt skip <int> ()}

specification of time series input files not to be taken into consideration, p.\,\pageref{CINDERELLA_keyword.skip}

\subsubsection{\tt target <int> (start index)}

specification of time series input files to be regarded as target datasets, p.\,\pageref{CINDERELLA_keyword.target}

\subsubsection{\tt tol <double> (0)}

{\sc Cinderella} frequency tolerance parameter, p.\,\pageref{CINDERELLA_keyword.tol}

\subsubsection{\tt transam:mean}

amplitude transformation using the mean observable for photon statistics, p.\,\pageref{CINDERELLA_keyword.transam:mean}

\subsubsection{\tt transam:none}

no amplitude transformation at all, p.\,\pageref{CINDERELLA_keyword.transam:none}

\subsubsection{\tt transam:ppsc}

amplitude transformation using the point-to-point scatter of the residual observable for photon statistics, p.\,\pageref{CINDERELLA_keyword.transam:ppsc}

\subsubsection{{\tt transam:rms} (default)}

amplitude transformation using the rms residual observable for photon statistics, p.\,\pageref{CINDERELLA_keyword.transam:rms}

\section{Online availability}

The ANSI-C code is available online at {\tt http://www.sigspec.org}. For further information, please contact P.~Reegen, {\tt peter.reegen@univie.ac.at}.

\acknowledgments{
PR received financial support from the Fonds zur F\"or\-de\-rung der wis\-sen\-schaft\-li\-chen Forschung (FWF, projects P14546-PHY, P17580-N2) and the BM:BWK (project COROT). Furthermore, it is a pleasure to thank M.~Gruberbauer (Univ.~of Vienna), D.\,B.~Guenther (St.~Mary's Univ., Halifax), M.~Hareter, D.~Huber, T.~Kallinger (Univ.~of Vienna), R.~Kusch\-nig (UBC, Vancouver), J.\,M. Matthews (UBC, Vancouver), A.\,F.\,J.~Moffat (Univ.~de Montreal), D.~Punz (Univ.~of Vienna), S.\,M. Rucinski (D.~Dunlap Obs., Toronto), D.~Sasselov (Harvard-Smithsonian Center, Cambridge, MA), L.~Schneider (Univ.~of Vienna), G.\,A.\,H.~Walker (UBC, Vancouver), W.\,W. Weiss, and K.~Zwintz (Univ.~of Vienna) for valuable discussion and support with extensive software tests. I acknowledge the anonymous referee for a detailed examination of both this publication and the corresponding software, as well as for the constructive feedback that helped to improve the overall quality a lot. Finally, I address my very special thanks to J.\,D.~Scargle for his valuable support.
}

\References{
Kallinger, T., Reegen, P., Weiss, W.\,W.~2008, A\&A, 481,  571\\

Reegen, P.~2005, in {\it The A-Star Puzzle}, Proceedings of IAUS 224, eds. J. Zverko, J. Ziznovsky, S.J. Adelman, W.W. Weiss (Cambridge: Cambridge Univ.~Press), p.~791\\

Reegen, P.~2007, A\&A, 467, 1353\\

Reegen, P.~2009, CoAst, submitted\\

Reegen, P., Gruberbauer, M., Schneider, L., Weiss, W.\,W.~2008, A\&A, 484, 601\\

Zwintz, K., Marconi, M., Kallinger, T., Weiss, W.\,W.~2004, in {\it The A-Star Puzzle}, Proceedings of IAUS 224, eds. J.~Zverko, J.~Ziznovsky, S.\,J.~Adelman, W.\,W.~Weiss (Cambridge: Cambridge Univ.~Press), p.\,353\\

Zwintz, K., Weiss, W.\,W.~2006, A\&A, 457, 237\\

}

%\TOCheadline{HELAS News}{8mm}{5.5mm}{-2mm}
%\input{caput_empty} 
%\input{caput_helas} 
%   \include{pH1_HELAS/HELAS_helasIII}

\setlength{\columnseprule}{0pt}\clearpage
\thispagestyle{empty}
\cleardoublepage
\thispagestyle{empty}

\end{document}